# Vectorized Cluster Search[*]

**Hans Gerd Evertz**

Supercomputer Computations Research Institute,
Florida State University,
Tallahassee, FL 32306
email: evertz@scri.fsu.edu

**Abstract**

Contrary to conventional wisdom, the construction of clusters on a lattice can easily be vectorized, namely over each "generation" in a breadth first search. This applies directly to e.g. the *single cluster* variant of the Swendsen-Wang algorithm. On a CRAY-YMP, total CPU time was reduced by a factor 3.5 – 7 in actual applications.

---

# Introduction

The vectorization described in this paper applies to "breadth first searches" in general. We shall describe it in the framework of cluster construction in a spin system. Monte Carlo simulations of many discrete and continuous spin systems have been revolutionized in recent years by the advent of "cluster algorithms" that eliminate or strongly reduce critical slowing down [1, 2, 3]. As an improvement over the original multi-cluster Swendsen-Wang algorithm [1], the *single-cluster variant* [2] further reduces critical slowing down for many systems. Where possible, it is now often the method of choice in computer time intensive simulations.

Until recently, there seemed to be one severe drawback to cluster algorithms, namely that they were supposed to be intrinsically non-vectorizable. The corresponding loss in speed often greatly reduced the original gain in critical slowing down. Recently the Swendsen-Wang (multi-cluster) algorithm has been put into vectorized [4] and into parallelized [5] form. In this paper we treat the more favourable *single-cluster-variant* and show how to vectorize the corresponding breadth-first-search.

## Cluster construction and breadth first search

We work on a euclidean 'square' lattice of arbitrary size and dimension. The abovementioned cluster algorithms amount to specifying a procedure for defining *bonds* (on/off) between lattice sites. Sets of lattice sites connected through such bonds are called clusters.

In the single-cluster algorithm, an initial site is chosen at random, and the cluster to which it belongs is then determined (constructed) by a search. (Bonds are often evaluated only during this search). Two commonly used search algorithms are "depth-first" and "breadth-first" search. We shall use the latter.

*Breadth first search:*

(1) start a list $C$ with one entry $i$, $i = i_0 =$ initial site
(2) for each neighbour $j$ of site $i$ that does not yet belong to the cluster:
  – determine if bond $<ij>$ is *on* or *off*;
  – if bond is *on*, then:
    add $j$ to list $C$ of cluster members;
    mark site $j$ as belonging to cluster
(3) repeat (2) for $i =$ next entry in list $C$, until list is exhausted.

## Vectorized search

The above search contains "generations" of sites, where the first generation is $\{i_0\}$, and each following generation consists of those direct neighbours of the previous generation that are newly entered into the list $C$.

We can now just *vectorize over each generation* in order to create the next one. There is one possible vector conflict, namely that a new site could be neighbour of more than one site of the current generation, and could thus be added to the list more than once. This conflict is easily avoided by considering only neighbours in *one direction* during each vectorized loop, and then treating directions in an outer loop. Thus the vectorized cluster search, as it runs on a CRAY-YMP, is this:

```
** Initialization
      ...define array:             neighbour_site(site,direction)
      ...define array or function: bond_is_on(site_1,site_2)
      ...initialize array:         site_is_in_cluster(site)=.false.
      ......
      list_end=1
      list_entry(list_end) = initial_site
      end_of_generation=0

** Cluster construction
   10 start_of_generation = end_of_generation + 1
      end_of_generation   = list_end
      do 20 mu=1,number_of_directions
CDIR$ IVDEP  ! tell compiler to ignore vector dependencies in "do 30"
         do 30 index=start_of_generation,end_of_generation
            site_i=list_entry(index)
            site_j=neighbour_site(site_i,mu)
            if(site_is_in_cluster(site_j)) goto 20
            if(bond_is_on(site_i,site_j)) then
               site_is_in_cluster(site_j) = .true.
               list_end = list_end+1
               list_entry(list_end) = site_j
            endif
   30    continue
   20 continue
      if(list_end.gt.end_of_generation) goto 10
```

Now the array "list_entry" contains a unique list of addresses of all sites in the cluster.

## Results

The vector length during a cluster search is initially 1. It will then rise one or more times and in the end be small again. Maximum and average length are influenced by cluster size, cluster shape, and dimensionality of the lattice. A small additional gain in speed (about 10 – 30 % on a CRAY) can be obtained be letting small loops run in scalar mode (below a length of about 4 on the CRAY-YMP).

Our vectorization is not suitable for the multi-cluster Swendsen-Wang algorithm, because the average cluster size is very small there.

We have been using the vectorized algorithm in several large scale Monte Carlo simulations on a CRAY-YMP. For a two-dimensional spin model [6] at an average cluster size of 1000, the observed average vector length was 34, and the complete update routine ran about 3.5 times faster than the non-vectorized version. A three-dimensional simulation [7] with complicated inner loop (complicated function "bond_is_on") ran about 7 times faster at average cluster size 1000. A similar gain was observed for a four-dimensional $O(4)$-model.

We have thus shown how a rather small modification will *vectorize the breadth-first-search algorithm*, resulting in a large gain in CPU time.